\documentclass{emulateapj}

\newcommand{\expnt}[2]{\ensuremath{#1 \times 10^{#2}}}   
\newcommand{\gsim}{\gtrsim}

\newcommand{\xmm}{\textit{XMM-Newton}}
\newcommand{\hr}{\ensuremath{^{\rm h}}}
\newcommand{\mn}{\ensuremath{^{\rm m}}}

\newcommand{\rxjk}{RX~J0720.4\ensuremath{-}3125}
\newcommand{\cxo}{\textit{CXO}}
\newcommand{\chandra}{\textit{Chandra}}
\newcommand{\rosat}{\textit{ROSAT}}
\newcommand{\rbs}{RBS~1223}
\newcommand{\rxj}{RX~J1308.6+2127}
\newcommand{\rxs}{1RXS~J130848.6+212708}

\newcommand{\secsec}{\ensuremath{\mbox{s s}^{-1}}}
\newcommand{\Hzsec}{\ensuremath{\mbox{Hz s}^{-1}}}

\newcommand{\fdot}{\ensuremath{\dot \nu}}
\newcommand{\fdotdot}{\ensuremath{\ddot \nu}}

\shorttitle{A Timing Solution for RX~J1308.6+2127}
\shortauthors{Kaplan \&\ van~Kerkwijk}

\slugcomment{ApJ Letters in Press}

\begin{document}

\title{A Coherent Timing Solution for the Nearby Isolated Neutron Star
RX~J1308.6+2127/RBS~1223}

\author{D.~L.~Kaplan\altaffilmark{1} and M.~H.~van Kerkwijk\altaffilmark{2}}

\altaffiltext{1}{Pappalardo Fellow; MIT Kavli Institute for Astrophysics and Space
  Research, Massachusetts Institute of Technology, 77 Massachusetts
  Avenue, 37-664D, Cambridge, MA 02139, USA; dlk@space.mit.edu}
\altaffiltext{2}{Department of Astronomy and Astrophysics, University
  of Toronto, 60 St.\ George Street, Toronto, ON M5S 3H8, Canada;
mhvk@astro.utoronto.ca}

\begin{abstract}
We present a phase-connected timing solution for the nearby isolated
neutron star \rxj\ (\rbs).  From dedicated \chandra\ observations as
well as archival \chandra\ and \xmm\ data spanning a period of five
years, we demonstrate that the 10.31-sec pulsations are slowing down
steadily at a rate of $\dot P=\expnt{1.120(3)}{-13}\mbox{ s s}^{-1}$.
Under the assumption that this is due to magnetic dipole torques, we
infer a characteristic age of 1.5~Myr and a magnetic field strength of
$\expnt{3.4}{13}$~G.  As with \rxjk, the only other radio-quiet
thermally emitting isolated neutron star for which a timing solution
has been derived, the field strength is roughly consistent with what was
inferred earlier from the presence of a strong absorption feature in
its X-ray spectrum.  Furthermore, for both sources the
characteristic age is in excess of the cooling age inferred from
standard cooling models.  The sources differ, however, in their timing
noise: while \rxjk\ showed considerable timing noise, \rxj\ appears
relatively stable.
\end{abstract}


\section{Introduction}

The sample of nearby, radio-quiet, isolated neutron stars discovered
by {\em ROSAT} (for a review, e.g., \citealt{haberl04}) is of
particular interest because of the unambiguous presence of strong,
broad absorption features in the X-ray spectra of most sources.  Since
the spectra appear thermal, these features almost certainly arise in
the neutron-star atmospheres.  The features have usually been
interpreted under the assumption of a pure hydrogen composition (not
unlikely, given the short settling times for neutron stars) and a
strong magnetic field (as suggested by the long, 3--10~s spin periods;
\citealt{hh98b}), with the absorption reflecting either the proton
cyclotron line or transitions between bound states of neutral hydrogen
(e.g., \citealt{hsh+03,vkkd+04}).

In order to help determine the nature of the absorption features, as
well as to elucidate what sets the isolated neutron stars apart from
young rotation-powered pulsars, we have started a program to obtain
phase-connected timing solutions, and use these to estimate ages and
magnetic field strengths.  Earlier, we presented our first results,
for \rxjk\ \citep{kvk05}; here, we consider a second source, \rxj.

\rxj\ (also known as \rbs\ and \rxs) was identified as a possible
nearby isolated neutron star by \citet{shs+99}.  The identification
was confirmed with the detection of a 5.16-s X-ray periodicity
\citep{hhss02} and a very faint ($V\approx 28$~mag) probable optical
counterpart with no  radio emission \citep*{kkvk02}.  By
comparing the periods measured from \chandra\ and archival \rosat\
data, \citet{hhss02} inferred a spin-down rate of $\dot
P=\expnt{(0.7-2.0)}{-11}\mbox{ s s}^{-1}$, implying a very strong
magnetic field of $\gsim 10^{14}$~G.  \citet{haberl04}, however,
showed that the 5.16-s periodicity was in fact the first harmonic,
implying a true period of 10.31~s (see also \citealt{hsh+03}), and
that the spin-down rate inferred earlier was likely erroneous.  This
was confirmed by \citet{shhm05}, who attempted to determine a
phase-connected timing solution, but failed due to cycle-count
ambiguities.

Here, we show that with additional \chandra\ observations specifically
obtained for timing purposes, we can obtain an unambiguous timing
solution.  We describe our analysis of the \chandra\ data, as well as
of archival \rosat, \chandra, and \xmm\ data, in \S~\ref{sec:obs}, and
use these to obtain a timing solution in \S~\ref{sec:timing}.  Since
our analysis closely follows the one used for \rxjk\ \citep{kvk05}, we
focus primarily on those aspects of the analysis that differ.  We
discuss the implications of our result in \S~\ref{sec:discuss}.

\section{Observations}\label{sec:obs}

Our primary data are eight observations taken with the Advanced CCD Imaging
Spectrometer \citep[ACIS;][]{gbf+03} aboard the \textit{Chandra X-ray
Observatory} (\textit{CXO}).  These were designed for timing
accuracy, consisting of two sets of four exposures in the
Continuous-Clocking (CC) mode geometrically spaced over a period of
about two weeks and separated by about half a year.  We combined these
with data from other \chandra\ observations, as well as from
observations with \textit{XMM-Newton} and \rosat.  A log of all observations\footnote{We do not use
data from \rosat\ observation 703848, since we, like \citet{hhss02} and
\citet{shhm05}, found there were insufficient counts to extract a
period or an arrival time.}  is given in Table~\ref{tab:obs}.

\begin{deluxetable*}{lrcrrl}
\tablewidth{0pt}
\tablecaption{Log of Observations and Times of Arrival\label{tab:obs}}
\tablehead{
\colhead{Instrument\tablenotemark{a}}&
\colhead{ID\tablenotemark{b}}&
\colhead{Date}&
\colhead{Exp.}&
\colhead{Counts}&
\colhead{TOA\tablenotemark{c}}\\
&&&\colhead{(ks)}&& \colhead{(MJD)}\\[-2.2ex]
}
\startdata
HRI          \dotfill& 704082 & 1998~Oct~01 &  4.8 &   498& 50824.2143496(30)\\
ACIS   $1/8$ \dotfill&    731 & 2000~Jun~24 &  9.5 &  7395& 51719.5182790(5) \\
PN/sw        \dotfill& 377-U2 & 2001~Dec~31 & 18.0 & 10633& 52274.2594926(10)\\
PN/ff        \dotfill& 561-S5 & 2003~Jan~01 & 27.0 & 66219& 52640.4325929(5) \\
MOS1         \dotfill& 561-S3 & 2003~Jan~01 & 29.0 & 14469& 52640.4265055(9) \\
MOS2         \dotfill& 561-S4 & 2003~Jan~01 & 29.0 & 14925& 52640.4265042(7) \\
PN/ff        \dotfill& 743-S3 & 2003~Dec~30 & 30.0 & 74587& 53003.4668264(6) \\
MOS1         \dotfill& 743-S1 & 2003~Dec~30 & 32.0 & 16086& 53003.4607372(7) \\
MOS2         \dotfill& 743-S2 & 2003~Dec~30 & 32.0 & 16523& 53003.4607388(9) \\
HRC          \dotfill&   4595 & 2004~Mar~30 & 90.1 & 33823& 53095.3847872(16)\\
ACIS CC      \dotfill&   5522 & 2005~Feb~14 & 16.0 &  8953& 53415.6878535(10)\\
              &   5523 & 2005~Feb~15 &  5.7 &  3254& 53416.5968822(17)\\ 
              &   5524 & 2005~Feb~19 &  5.2 &  2922& 53420.1717683(14)\\
              &   5525 & 2005~Mar~10 &  5.6 &  2923& 53439.0475044(11)\\
ACIS CC      \dotfill&   5526 & 2005~Jul~09 & 15.1 &  7684& 53560.2591584(8) \\
              &   5527 & 2005~Jul~10 &  5.1 &  2876& 53561.2549594(17)\\
              &   5528 & 2005~Jul~14 &  5.2 &  2937& 53565.7697939(18)\\ 
              &   5529 & 2005~Jul~29 &  5.2 &  3048& 53580.7839447(14)\\
\enddata
\tablenotetext{a}{HRI: High-Resolution Imager \citep{zdhk95} aboard \rosat.
  PN: \xmm's European Photon Imaging Camera with PN detectors
    \citep{sbd+01}, used in full-frame (ff) or small window (sw)
    mode, with thin filter.
  MOS1/2: European Photon Imaging Cameras with MOS detectors
    aboard \xmm\ \citep{taa+01}, used in small-window mode with thin
    filter.
  HRC: High-Resolution Camera for spectroscopy aboard \chandra\ (HRC-S;
    \citealt{kck+97}), used with the Low-Energy Transmission Grating (LETG).
  ACIS: \chandra's Advanced CCD Imaging Spectrometer,
    with the S3 chip used either in 1/8-subarray  or in Continuous
    Clocking (CC) mode.} 
\tablenotetext{b}{Observation identifier (\cxo, \rosat) or revolution
  number and exposure identifier (\xmm).}
\tablenotetext{c}{The TOA is defined as the time of maximum light of
  the pulsation after the deepest minimum
  closest to the middle of each observation, and is given with
  1-$\sigma$ uncertainties.}
\end{deluxetable*}

For the \chandra\ data, we processed the level-1 event lists to the
level-2 stage following standard procedures and the latest calibration
set (\texttt{CALDB} version 3.1.0).  For the ACIS data, we extracted
events within $1\arcsec$ of the source with energies between 0.2
and 2.0~keV, and then applied a clock correction of $284.7\,\,\mu{\rm
s}$ \citep*{dhm03}.  For the HRC-S/LETG data, we extracted
zeroth-order events from a circle with radius of $1\farcs3$, and
first-order events using the standard LETG spectral extraction
windows, but limited to $10\,\mbox{\AA}\leq\lambda\leq65\,\mbox{\AA}$.
Finally, we used the \texttt{axbary} program to barycenter all of the
events (using the X-ray position: $\alpha_{\rm
J2000}=13\hr08\mn48\farcs27$, $\delta_{\rm
J2000}=+21\degr27\arcmin06\farcs8$; \citealt{kkvk02})

For the \xmm\ data, we used the standard procedures \texttt{emchain}
and \texttt{epchain} (\texttt{XMMSAS} version 6.5.0) to reprocess the
observations.  Next, we extracted events within 37\farcs5 of the
source position (using standard quality and pattern selections) with
energies from 0.12~keV to 1.2~keV, and used \texttt{barycen} to
convert the arrival times to the solar-system barycenter.  And finally,
for the \rosat\ HRI data, we extracted the events within $12\farcs5$
of the source.  We barycentered these using the \texttt{FTOOLS}
programs \texttt{ABC} and \texttt{BCT}, and converted the event times
from Coordinated Universal Time (UTC) to Barycentric Dynamical Time
(TDB) using the corrections supplied in \citet[][p.\ 14]{allen}.

\section{Timing Analysis}\label{sec:timing}

Our goal is to use times-of-arrival (TOAs) to infer a phase-coherent
timing solution in which each cycle of the source is accounted for.
To measure TOAs, we need an initial reference period.  We determined
this from the ACIS CC data-sets using a $Z_2^2$ test \citep{bbb+83},
which combines power from the 10.31-s fundamental with that from the
5.16-s  harmonic (we could not detect significant power in any
higher harmonics).  We calculated $Z_2^2$ both for the individual sets
of four observations (observations 5522--5525 and 5526--5529) as well
as for the eight observations combined.  The overall $Z_2^2$ spectrum
has a single, well-defined peak, and all possible aliases are at
considerably lower significances.  The measured period is
$P=10.31252293(9)$~s (here and below, numbers in parentheses indicate
the formal 1-$\sigma$ uncertainties in the last digit unless otherwise
indicated).  

\begin{figure}
\plotone{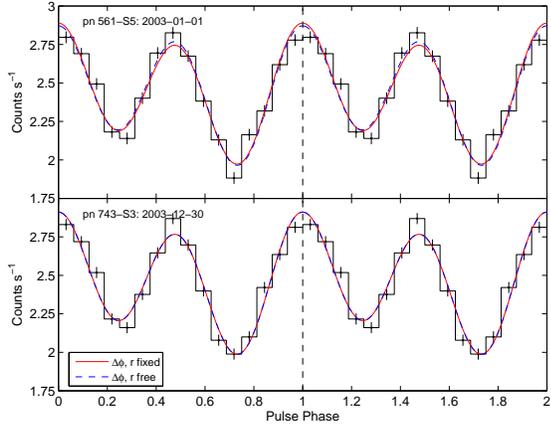}
\caption{Pulse profiles of the two long EPIC-PN observations, showing
  the different fits to the binned lightcurves.  Top: data from
  revolution 561; bottom: data from revolution 743.  The solid red
  lines show the fits using Eqn.~\ref{eqn:fitfh} with $\Delta\phi_2$
  and $r_2$ free , while the dashed blue lines (largely
  indistinguishable from the red lines) show the fits with
  $\Delta\phi_2$ and $r_2$ fixed to the average best-fit  values of
  0.03 and 2.8.  The zero point in phase has been chosen such that it
  coincides with the maximum following the deepest minimum, which we
  define as the reference position for the TOAs; it is 
  indicated by the vertical dashed line at phase 0.}
\label{fig:pn}
\end{figure}

Using the period derived above, we constructed binned light curves
(with 16 phase bins) for all of the observations.  Unlike \rxjk, \rxj\
has non-sinusoidal pulsations and hence determining TOAs by fitting a
single sinusoid would be inappropriate.  Instead we fit for both the
first harmonic and fundamental, with a possible phase shift between
them:
\begin{equation}
N_{i}  =  A\left[ \sin\left(2\pi(f_0 t_i + \phi_0)\right)
       + r_2\sin\left(4\pi(f_0 t_i + \phi_0+\Delta \phi_2)\right)\right]+C,
\label{eqn:fitfh}
\end{equation}
where $N_{i}$ is the number of counts in bin $i$, $f_0=0.096969481$~Hz
is the frequency of the fundamental from above, $t_i$  the time
of bin $i$ on the interval $[0,P)$, and the parameters are amplitude
$A$, phase $\phi_0$, amplitude ratio $r_2$, phase offset $\Delta
\phi_2$, and constant offset $C$.  As can be seen in
Figure~\ref{fig:pn}, this model provides a good fit to even the
highest-quality lightcurves, which are those derived from the long
EPIC-PN observations.  The inferred phase offset and relative
amplitude are similar for both observations, with $\Delta
\phi_2\approx 0.03$, and $r_2\approx 2.8$.

Assuming that the pulse profile is not varying, we measured arrival
times for all of the data using the above model, but with
$\Delta\phi_2$ and $r_2$ fixed at 0.03 and 2.8, respectively, which
generally provides a good fit to the light curves (see
Fig.~\ref{fig:pn} for the EPIC-PN data).  Our resulting arrival times
are listed in Table~\ref{tab:obs}.  Here, we assigned these times to
the time of the maximum of the model lightcurve that follows the
lowest minimum and is nearest to the middle of the observation (see
Fig.~\ref{fig:pn}; this is the spectrally harder maximum from
\citealt{shhm05}).  For completeness, we note that the TOAs derived
using the above model are consistent with those one would find using a
simpler model consisting of only the first harmonic (i.e., a sinusoid
at $P=5.16$~s): the mean absolute difference is $0.2\sigma$.

We determined a timing solution from the TOAs using an iterative
procedure, as in \citet{kvk05}.  We started with the \chandra\ ACIS CC
TOAs and found that we were able to fit these without the need for a
frequency derivative and with no cycle ambiguity, consistent with the
power spectrum analysis.  However, between the first ACIS CC TOA and
the preceding TOA (\chandra\ HRC) there is a gap of 320~days and we
found that a constant-frequency model led to a poor fit
($\chi^2=202.7$ for 7 degrees of freedom).  Including a frequency
derivative $\fdot$, the cycle count over the gap became ambiguous by
$\pm 1$ cycle.  The three possibilities---2683550, 2683551, and
2683552 cycles---lead to three possible solutions for $\fdot$:
$\expnt{2.1}{-15}\mbox{ Hz s}^{-1}$, $\expnt{1.0}{-15}\mbox{ Hz
s}^{-1}$, and $\expnt{-1.1}{-15}\mbox{ Hz s}^{-1}$ .  Fortunately, the
addition of the remaining data eliminates the first two of those
solutions: with the \xmm\ TOAs (and the remaining ACIS point), they
have $\chi^2=439.1$ and 1107.6 with TOA rms of 0.041~s and 0.099~s,
respectively, while the third has $\chi^2=18.7$ with a TOA rms of 0.010~s
(all for 14 degrees of freedom).

The timing solution derived from the \chandra\ and \xmm\ TOAs is
presented in Table~\ref{tab:ephem}.  As indicated by the value of
$\chi^2$, the data are well reproduced by simple spin-down; the
addition of a cubic (\fdotdot) term leads to only an insignificant
reduction in $\chi^2$, from 18.7 to 18.6, and the inferred value of
$\fdotdot=\expnt{-7(15)}{-26}\mbox{ Hz s}^{-2}$ is consistent with
zero.

\begin{deluxetable}{p{1.5in} c}[t]
\tablewidth{0pt}
\tablecaption{Measured and Derived Timing Parameters for \rxj\ From
  \chandra\ and \xmm\ Data\label{tab:ephem}}
\tablehead{
\colhead{Quantity\tablenotemark{a}} & \colhead{Value}
\\[-2.2ex]
}
\startdata
Dates (MJD) \dotfill         & 51720--53581 \\
$t_{0}$ (MJD)\dotfill        & 53415.687853(2) \\
$\nu$ (Hz) \dotfill          & 0.0969694896(2)  \\
$\dot \nu$ (\Hzsec) \dotfill & $\expnt{-1.053(3)}{-15}$ \\
TOA rms (s) \dotfill         & 0.010 \\
$\chi^2$/DOF \dotfill        & 18.7/14=1.34 \\[1ex]
$P$ (s)\dotfill              & 10.31252206(2) \\
$\dot P$ (\secsec)\dotfill   & $\expnt{1.120(3)}{-13}$ \\
$\dot E$ ($\mbox{erg s}^{-1}$)\dotfill& $\expnt{4.0}{30}$ \\
$B_{\rm dip}$ (G) \dotfill   & $\expnt{3.4}{13}$ \\
$\tau_{\rm char}$ (yr)\dotfill& $\expnt{1.5}{6}$ \\
\enddata
\tablecomments{Uncertainties quoted are twice the formal 1-$\sigma$
  uncertainties in the fit.}
\tablenotetext{a}{$\tau_{\rm char}=P/2{\dot P}$ is the characteristic age,
  assuming an initial spin period $P_0\ll P$ and a constant magnetic
  field; $B_{\rm dip}=\expnt{3.2}{19}\sqrt{P{\dot P}}$ is the
  magnetic field inferred assuming spin-down by dipole radiation;
  $\dot E=10^{45}I_{45}4\pi^2\nu\dot\nu$ is the
  spin-down luminosity (with $I=10^{45}I_{45}\,\,{\rm g}\,\,{\rm
    cm}^{2}$ the moment of inertia).} 
\end{deluxetable}

\begin{figure}
\plotone{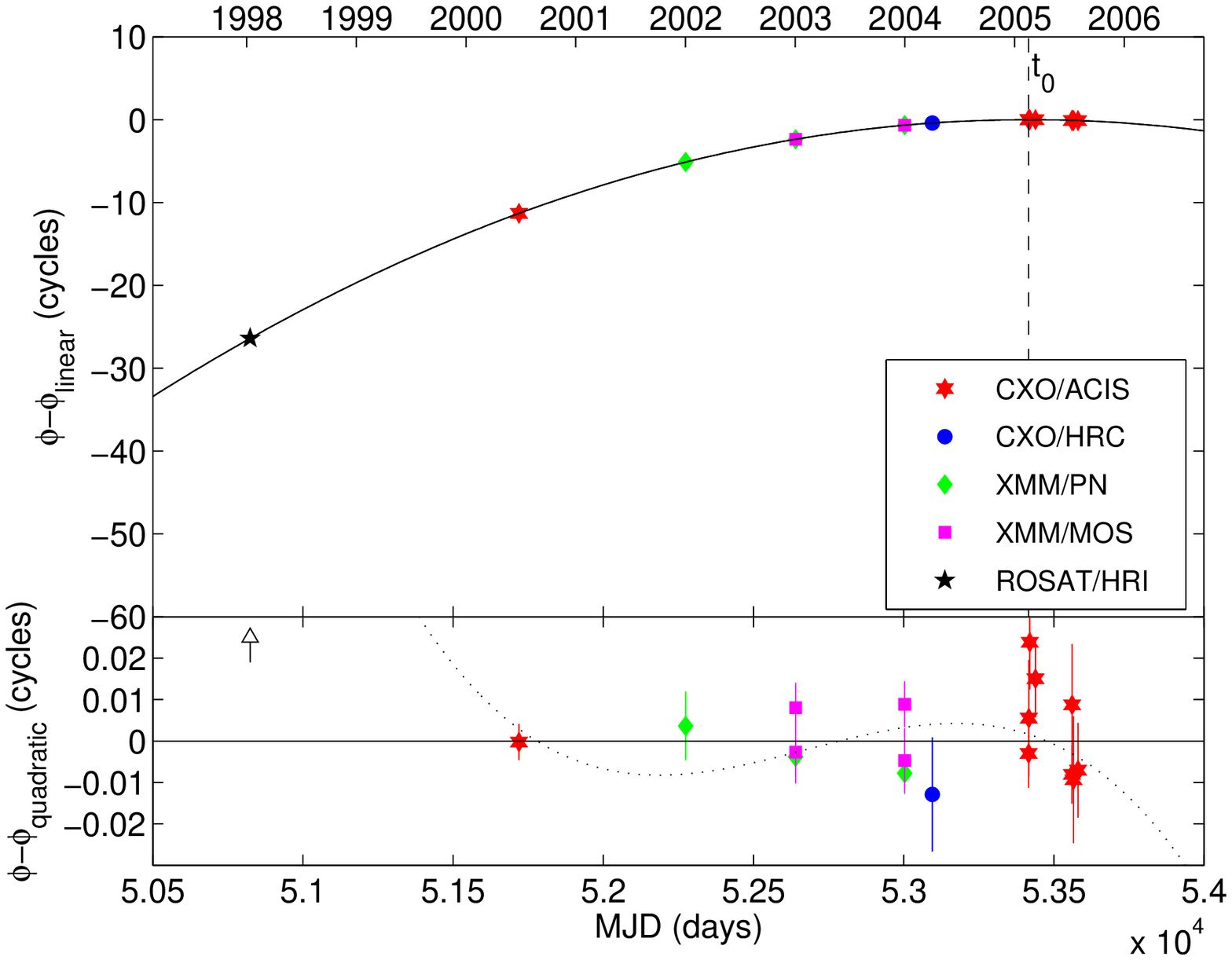}
\caption{Phase residuals for \rxj.  The top panel shows the residuals
  for each TOA compared to a linear ($\fdot=0$) model.  The solid
  curve gives the best-fit quadratic ($\fdot \neq 0$, $\fdotdot=0$)
  ephemeris for all the \chandra\ and \xmm\ data
  (Tab.~\ref{tab:ephem}).  The vertical dashed line indicates the
  reference time $t_0$.  The bottom panel shows the residuals relative
  to the quadratic model; the \rosat\ HRI point is off the scale at
  0.14~cycles, and so is indicated with an arrow in this panel.  We
  also show a best-fit cubic model fitted to all of the data including
  the \rosat\ point ($\fdotdot \neq 0$; dotted line). }
\label{fig:resid}
\end{figure}

Unlike the \chandra\ and \xmm\ data, however, the \rosat\ HRI point is
slightly discrepant from the fit in Table~\ref{tab:ephem}, exceeding
it by 0.15(3)~cycles.  If we include it in the fit, the overall
solution does not change drastically ($\fdot$ becomes
$\expnt{-1.050(2)}{-15}\mbox{ Hz s}^{-1}$) and the deviation does
decrease to $0.11(3)$~cycles, but the overall $\chi^2$ increases to
$43.5$ for 15 degrees of freedom (with a TOA rms of 0.0271~s).
The deviation could be intrinsic (a glitch, timing noise, precession,
changes in pulse profile), but might also be instrumental, perhaps
related to differences in energy responses between the different
instruments.  Unfortunately, 
the sampling is too sparse to determine a
unique solution.  As an example, we show in Figure~\ref{fig:resid} a
cubic fit 
that matches the HRI point reasonably
well, with $\chi^2=22.5$ for 14 degrees of freedom, TOA rms of
0.011~s, and $\fdotdot=\expnt{-2.5(5)}{-25}\mbox{ Hz s}^{-2}$.  We
stress, however, that, in essence, this solution approximates the
\chandra\ and \xmm\ data with a $\fdotdot=0$ model and then adjusts
$\fdotdot$ to pass through the HRI point.  Hence, it does not gives
much additional insight.

\section{Discussion \& Conclusions}\label{sec:discuss}

The spin period and spin-down rate we derive for \rxj\ are quite
similar to those of \rxjk, the only other isolated neutron star with a
timing solution \citep{kvk05}, placing both of them above the pulsar
``death-line'' in a $P$-$\dot P$ diagram despite their lack of radio
emission \citep{kkvk02,kvkm+03}.  Hence, assuming both spin down by
magnetic dipole radiation, \rxj\ and \rxjk\ have similar inferred
magnetic field strengths ($B_{\rm dip}=3.4$ and $\expnt{2.4}{13}$~G,
respectively), characteristic ages ($\tau_{\rm char}=1.5$ and
1.9~Myr), and spin-down luminosities ($\dot E=4.0$ and
$\expnt{4.7}{30}\mbox{ ergs s}^{-1}$).  Both sources also have thermal
X-ray spectra with temperatures of $\sim\!1\times10^6~$K, but
superposed on these are rather different absorption features: in \rxj,
the absorption feature is centered near or below 300~eV and is very
wide and strong, with an equivalent width of $>\!150$~eV
\citep{hsh+03}; in \rxjk, on the other hand, no absorption was
apparent until 2004, when an absorption feature with only a slightly
higher energy but much smaller equivalent width ($\sim\!40$~eV)
appeared \citep{hztb04,dvvmv04}.  If the magnetic fields, line
energies, and blackbody temperatures are all similar, what would
account for the significant difference in line strength?

The interpretation of the absorption lines is still a matter of
debate.  For \rxj, \citet{hsh+03} suggested proton cyclotron
absorption in a field of (2--6)$\times 10^{13}$~G, which could match both
the energy and strength of the observed feature (for a gravitational
redshift of $z=0.3$, the observed
line energy would be at
100--300~eV).  As the feature in \rxjk\ was so much weaker but at a
similar energy, \citet{vkkd+04} argued that for that source the proton
cyclotron line was below the observed band and that the absorption
was due to the $0\rightarrow2$ transition between tightly-bound states
of neutral hydrogen in a $\sim\!\expnt{2}{13}$~G field.

Qualitatively, the field strengths inferred from our timing
measurements agree with the above expectations: the field of \rxj\ is
stronger than that of \rxjk, while a weaker field would be expected if
the absorption in both sources were due to the same mechanism.
Quantitatively, a larger difference between the two sources was
expected.  However, this discrepancy may simply reflect 
orientation and/or substructure in the magnetic field---higher-order
multipoles near the surface would affect the emission properties but
not the spin-down rate.  Such effects may also be responsible for the
difference in pulse profile: double-peaked for \rxj, and sinusoidal
for \rxjk.  Phase-resolved X-ray spectroscopy coupled with
observations of more sources are probably the best ways to disentangle
these effects, and efforts are underway
\citep[e.g.,][]{hztb04,shhm05}.

Another open issue is that the characteristic ages of \rxjk\ and now
\rxj\ are three to four times larger than the values of
$\sim\!0.5$~Myr one would expect for simple cooling models
\citep{hh98b,kkvkm02,zhc+02} and, in the case of \rxjk, tracing the object
back along its trajectory 
to a likely birth
location (\citealt*{mzh03}, \citealt{kaplan04}).  Indeed, this age
discrepancy may extend to the other isolated neutron stars discovered
by \rosat: all have similar temperatures and
thus likely similar cooling ages, and most also have similar periods
(3--10~s) and, based on the similar energies at which they show X-ray
absorption features (0.3--0.7 keV), similar magnetic field strengths,
implying similar characteristic ages.


We first consider whether the discrepancy could result from the
characteristic age being an overestimate.  In general, for a spin-down
torque $\propto\nu^n$, the pulsar's spin-down age is given by
$t_{\rm sd} = \left[P/(n-1)\dot P\right]\left[1-\left(P_0/P\right)^{n-1}\right]$,
where $P_0$ is the initial spin period and $n=\nu\ddot\nu/\dot\nu^2$
is the ``braking index,'' equal to 3 under the assumption of magnetic
dipole radiation \citep[e.g.,][p.\ 111]{mt77}.  For $P_0\ll P$ and
$n=3$, one recovers the characteristic age $\tau_{\rm char}\equiv P/2\dot P$, but
if $P_0$ is not much smaller than the current period~$P$, the
characteristic age is an overestimate.  While we cannot exclude the
required birth periods of 7--8~s, 
there is as yet no concrete indication that neutron stars are born
with periods longer than 100~ms \citep[e.g.,][]{kh02,ghs05}.

For \rxjk, this led us to discuss the possibility that the neutron
star formed in a binary system and accreted matter before being
ejected in a second supernova; at ejection, the neutron star would
still be hot from the accretion, but spinning slowly \citep{kvk05}.  A
variation on this model would involve accretion from a residual debris
disk such as that recently discovered around the anomalous X-ray
pulsar 4U~0142+61 \citep*{wck05}, but this assumes that the accretion
disk persists for a sufficient time and can affect the spin-down of
the neutron stars, both of which are far from clear.  In either case,
if accretion played a role, it might explain why the
spectral properties of the isolated neutron stars are rather different
from those of the radio pulsar population.

Unfortunately, the above solution for the age discrepancy is not
unique.  First, it is possible that the spin-down was not due to a
constant dipole but that the magnetic field decayed (effectively, this
implies $n>3$).  Second, the cooling age could be
incorrect due to non-standard cooling.  For instance, the energy
released by magnetic field decay might keep a neutron star hotter
(\citealt{hk98}; note, however, that these authors do not expect
significant effects for the field strengths we infer) and longer
cooling times are also expected for a light neutron star
\citep{ygk+04}.  A prolonged cooling timescale seems somewhat
unlikely in light of the kinematic age estimates, though it should be
kept in mind that these may also not be unique: for \rxjk, several
different birth places are possible \citep{mzh03,kaplan04}.

Finally, when fitted with just a simple spin-down model, \rxjk\ showed
significant timing residuals of 0.31~s (root-mean-square), much larger
than the uncertainties on the TOAs.  In contrast, for \rxj\ the
root-mean-square residuals are only 0.010~s (excluding the HRI point),
consistent with measurement errors.  Since the sources were observed
with the same instruments, and have similar spectral shapes, the
differences in timing behaviors are likely not instrumental.  Instead,
they may well reflect the fact that both the spectrum and pulse
profile of \rxjk\ are varying with time \citep{dvvmv04,vdvmv04}, while
\rxj\ appears to be stable.

\acknowledgements We acknowledge support through Chandra grant
GO5-6050A.

\bibliographystyle{apj}

\end{document}